\documentclass[manuscript]{acmart}
\AtBeginDocument{%
  \providecommand\BibTeX{{%
    \normalfont B\kern-0.5em{\scshape i\kern-0.25em b}\kern-0.8em\TeX}}}

\setcopyright{none}

\acmConference[FaccTRec@RecSys '23]{17th ACM Conference on Recommender Systems}{Sep.~18--22, 2023}{Singapore, Singapore}
\usepackage{color}
\usepackage{hyperref}
\usepackage{xspace}
\usepackage{graphicx}
\usepackage{acronym}
\usepackage{multicol}
\usepackage{multirow}
\usepackage{dirtree}
\usepackage{booktabs}
\usepackage{adjustbox}
\usepackage{enumitem}
\usepackage{amsthm}
\usepackage{subfig}
\usepackage{amsmath}
\usepackage{algorithm}
\usepackage{algpseudocode}

\acrodef{RS}{Recommender System}
\acrodef{POI}{Point-of-Interest}
\acrodef{CF}{Collaborative Filtering}
\acrodef{GUI}{Graphical User Interface}

\newcommand{\partitle}[1]{\vspace{2mm}\noindent\textbf{#1}}

\begin{document}

\title{Provider Fairness and Beyond-Accuracy Trade-offs in Recommender Systems}


\author{Saeedeh Karimi}
\affiliation{%
  \institution{University of Zanjan}
  \country{Iran}}
\email{karimi.saeedeh@znu.ac.ir}

\author{Hossein A.~Rahmani}
\affiliation{%
  \institution{University College London}
  \country{UK}}
\email{hossein.rahmani.22@ucl.ac.uk}

\author{Mohammadmehdi Naghiaei}
\affiliation{%
  \institution{University of Southern California}
  \country{US}}
\email{naghiaei@usc.edu}

\author{Leila Safari}
\affiliation{%
  \institution{University of Zanjan}
  \country{Iran}}
\email{lsafari@znu.ac.ir}

\renewcommand{\shortauthors}{S.~Karimi, H.~A.~Rahmani, M.~Naghiaei, L.~Safari}

\begin{abstract}
Recommender systems, while transformative in online user experiences, have raised concerns over potential provider-side fairness issues. These systems may inadvertently favor popular items, thereby marginalizing less popular ones and compromising provider fairness. While previous research has recognized provider-side fairness issues, the investigation into how these biases affect beyond-accuracy aspects of recommendation systems—such as diversity, novelty, coverage, and serendipity—has been less emphasized. In this paper, we address this gap by introducing a simple yet effective post-processing re-ranking model that prioritizes provider fairness, while simultaneously maintaining user relevance and recommendation quality. We then conduct an in-depth evaluation of the model's impact on various aspects of recommendation quality across multiple datasets. Specifically, we apply the post-processing algorithm to four distinct recommendation models across four varied domain datasets, assessing the improvement in each metric, encompassing both accuracy and beyond-accuracy aspects. This comprehensive analysis allows us to gauge the effectiveness of our approach in mitigating provider biases. Our findings underscore the effectiveness of the adopted method in improving provider fairness and recommendation quality. They also provide valuable insights into the trade-offs involved in achieving fairness in recommender systems, contributing to a more nuanced understanding of this complex issue.
\end{abstract}





\maketitle

\section{Introduction and Context}
\label{sec:intro}
The rapid proliferation of digital content and the abundance of information available on the internet have made recommendation systems indispensable tools for users to discover relevant and engaging items. These systems are widely employed across various domains such as e-commerce, entertainment, and news, among others. Despite their widespread adoption, recommendation systems often suffer from popularity bias, a phenomenon where popular items are disproportionately recommended at the expense of less popular or long-tail items \cite{rahmani2022experiments,rahmani2022unfairness}. This leads to a skewed exposure of items and potentially unfair treatment of providers, particularly those offering niche or less popular content. Additionally, popularity bias may adversely affect the diversity, novelty, and serendipity of recommendations, thereby limiting users' experiences and the discovery of new content. In some cases, this bias could reinforce echo chambers or marginalize certain content providers, leading to broader societal implications~\cite{geschke2019triple,naghiaei2022,rahmani2022role}.

Previous research in the field of recommendation systems has primarily focused on improving accuracy, with little attention paid to the trade-offs between provider fairness and other essential dimensions of recommendation quality \cite{zhao2022investigating,naghiaei2022towards,rahmani2022exploring}. Some existing approaches to mitigate popularity bias include re-sampling techniques~\cite{li2019repair,naghiaei2022pycpfair} and diversity-aware algorithms \cite{schelenz2021diversity}. 
However, these methods often overlook the complex interplay between provider fairness and other important aspects of recommendation quality, such as diversity, serendipity, and novelty. This leaves a significant gap in our understanding of the broader impact of fairness-aware algorithms.  

In contrast, our research takes a comprehensive approach to examine the intricate and complex trade-offs that come into play when measures to enhance fairness in recommendation systems are implemented. We introduce a method, inspired by existing post-processing techniques~\cite{li2021user}, that is uniquely designed to promote fairness among providers without significantly sacrificing accuracy. 
Post-processing methods are particularly advantageous as they strike a balance between provider fairness, user relevance, and beyond-accuracy performance~\cite{naghiaei2022cpfair}. Furthermore, these methods are recommendation model agnostic, enhancing their suitability for real-world applications. Furthermore, we delve into an in-depth investigation of the consequences and trade-offs between provider fairness and recommendation quality across four recommendation algorithm baselines and four datasets in different domains. To thoroughly investigate the trade-offs between provider fairness and recommendation quality, we pose several research questions that guide our study:

\begin{enumerate}[label=\textbf{RQ\arabic*},leftmargin=*]
    \item How does the proposed post-processing re-ranking optimization framework affect the exposure of long-tail items and overall provider fairness in recommendation systems?
    \item What are the consequences of improving provider fairness on other important aspects of recommendation quality, such as diversity, serendipity, and novelty? 
    \item How do the trade-offs between provider fairness and recommendation quality manifest across different recommendation algorithms and real-world datasets? 
\end{enumerate}
Our experiments offer compelling evidence of the effectiveness of our proposed framework. It not only improves the exposure of long-tail items and enhances provider fairness, but also preserves other important aspects of recommendation quality such as diversity, serendipity, and novelty. The subsequent sections present a detailed discussion of our methodology, experiments, and findings.




\section{Proposed Provider Fairness Model}
\label{sec:model}

In this section, we propose a simple yet effective model to improve fairness among providers to analyse the effect of mitigating provider bias on beyond-accuracy metrics.

\partitle{Provider Fairness.}
In recommendation systems, provider fairness can be addressed by mitigating the popularity bias, where popular items are disproportionately recommended at the expense of less popular or long-tail items. Let $U = \{u_1, u_2, \dots, u_m\}$ denote the set of users and $I = \{i_1, i_2, \dots, i_n\}$ represent the set of items in the recommendation system, $m$ and $n$ are the number of users and items, respectively. We divide the items into two groups, \textit{popular} (or short-head) and \textit{non-popular} (or long-tail) items. To measure the average difference in the chosen popularity measure between popular and non-popular items across all users, we introduce binary decision variables $Y_{j}$, where $Y_{j} = 1$ if item $j$ is a popular item, and $Y_{j} = 0$ if item $j$ is a non-popular item. We define the provider fairness metric $F$ as follows:

\begin{equation}
F = \frac{1}{m} \sum_{i=1}^m \left(\sum_{j=1}^n Y_{j} * X_{ij} - \sum_{j=1}^n (1 - Y_{j}) * X_{ij}\right)
\end{equation}

The provider fairness metric, $F$, captures the difference in exposure between popular and non-popular items in the recommendation system. By considering this metric, our goal is to balance the exposure of items from both groups, ensuring that long-tail items receive adequate visibility while still delivering relevant recommendations to users. An $F$ of $0$ indicates equal exposure to long-tail and short-head items.

\partitle{Fairness-aware Algorithm.}
We define a framework that generates fairness-aware recommendation lists by applying a re-ranking method to the output of a baseline recommendation model. Our fairness-aware algorithm is further illustrated by the Algorithm \ref{alg:rerank}, which provides a step-by-step implementation of the re-ranking process. The algorithm takes as input the score matrix $R$, the number of items to recommend $K$, the binary popularity matrix $Y$, and the hyperparameter $\lambda$.  The score matrix $R$ is generated by baseline recommendation models and has dimensions $m \times n$, where $R_{ij}$ represents the predicted score or preference of user $i$ for item $j$. We introduce binary decision variables $X_{ij}$, where $X_{ij} = 1$ if item $j$ is selected in the re-ranked list for user $i$, and $X_{ij} = 0$ otherwise. Furthermore, to incorporate the provider fairness aspect into the re-ranking process, we introduce a hyperparameter lambda ($\lambda$) that controls the trade-off between recommendation score and provider fairness.  In the integer programming model, we aim to maximize the sum of scores while minimizing the difference in the chosen popularity measure between popular and non-popular items, controlled by the hyperparameter $\lambda$. The ReRank algorithm formulates and solves the integer programming problem to generate the re-ranked recommendation matrix $X$. 

\begin{algorithm}
\caption{Re-ranking Recommendations with Provider Fairness}
\label{alg:rerank}
\begin{algorithmic}[1]
\Procedure{ReRank}{$R$, $K$, $Y$, $\lambda$}
\State Initialize $m \times n$ binary matrix $X$
\State \textbf{Input:} Score matrix $R \in \mathbb{R}^{m \times n}$, number of items to recommend $K$, binary popularity matrix $Y \in \{0, 1\}^{n}$, hyperparameter $\lambda$
\State \textbf{Output:} Re-ranked recommendation matrix $X \in \{0, 1\}^{m \times n}$

\State Define the objective function as: $Z(X) = \sum_{i=1}^m \sum_{j=1}^n (R_{ij} * X_{ij}) - \lambda * F$

\State Define the constraints as:
\begin{itemize}
  \item $\sum_{j=1}^n X_{ij} = K, \quad \forall i \in \{1, \dots, m\}$
  \item $X_{ij} \in \{0, 1\}, \quad \forall i \in \{1, \dots, m\}, \forall j \in \{1, \dots, n\}$
  \item $Y_{ij} \in \{0, 1\}, \quad \forall j \in \{1, \dots, n\}$
\end{itemize}

\State Formulate the integer programming problem: $\max_{X} Z(X)$ subject to the constraints

\State Solve the integer programming problem using a solver (e.g., Gurobi or CPLEX)

\State Return the re-ranked recommendation matrix $X$
\EndProcedure
\end{algorithmic}
\end{algorithm}


The constraints of the model ensure that for each user $i$, exactly $K$ items are selected in the re-ranked list (i.e., $X_{ij} = 1$ for $K$ items). Integer programming is an NP-hard problem, which means that the runtime complexity increases exponentially with the size of the problem. However, modern integer programming solvers such as Gurobi\footnote{\url{https://www.gurobi.com/}} or CPLEX\footnote{\url{https://www.ibm.com/products/ilog-cplex-optimization-studio}} can efficiently handle problems of moderate size, making them suitable for solving the re-ranking problem in this context. In this formulation, the hyperparameter $\lambda$ allows for controlling the trade-off between maximizing the sum of recommendation scores and improving provider fairness. When $\lambda$ is set to a high value, the optimization model places more emphasis on provider fairness, while lower values of $\lambda$ prioritize recommendation scores. When $\lambda = 0$, the re-ranked list will be the same as the initial top-$N$ recommendations, as no emphasis is placed on provider fairness. By tuning $\lambda$, practitioners can balance the need for accurate recommendations with the importance of giving exposure to less popular items, thus creating a more diverse and fair recommendation experience for users and content providers.

\section{Experimental Setup}
\label{sec:experiments}
This section briefly describes the datasets, baseline models, and evaluation metrics. To foster the reproducibility of our experiments\footnote{We release our codes and dataset for the reproducibility and future work at \url{https://github.com/rahmanidashti/BeyondAccProvider}}, we implemented and evaluated all the recommendations with the open-source Python-based recommendation toolkit Cornac \cite{salah2020cornac}.

\subsection{Datasets}
To evaluate our proposed approach, we use four publicly available datasets from different domains and the details of these datasets are given in Table \ref{tbl:datasets}. We divided the dataset into three subsets: a training set (70\%), a validation set (10\%), and a test set (20\%), which were used for all model training and evaluation. Additionally, we classified items into two groups: \textit{popular} or \textit{short-head} items, representing the top 20\% of items with the most interactions, while the remaining items were classified as \textit{non-popular} or \textit{long-tail} or items.


\begin{table}
  \caption{Statistics of the datasets: $\left| U \right|$ is the number of users, $\left| I \right|$ is the number of items, $\left|P\right|$ is the number of interactions, $\left| SI \right|$ is the number of popular (short-head) items, $\left| LI \right|$ is the number of non-popular (long-tail) items.}
  \centering
  \label{tbl:datasets}
  \begin{adjustbox}{max width=\textwidth}
  \begin{tabular}{llllllllll}
    \toprule
    \textbf{Dataset} & $\left| U \right|$ & $\left| I \right|$ & $\left| P \right|$ & $\frac{\left| P \right|}{\left| U \right|}$ & $\frac{\left| P \right|}{\left| I \right|}$ & \%Sparsity & Domain & $\left| SI \right|$ & $\left| LI \right|$ \\
     
    \midrule
    \textbf{Epinions} & 2,677 & 2,060 & 103,567 & 38.6 & 50.2 & 98.12\% & Opinion & 412 & 1,648 \\
    \textbf{BookCrossing} & 1,136 & 1,019 & 20,522 & 18.0 & 20.1 & 98.22\% & Book & 203 & 816 \\
    \textbf{Gowalla} & 1,130 & 1,189 & 66,245 & 58.6 & 55.7 & 95.06\% & POI & 237 & 952 \\
    \textbf{Last.fm} & 1,797 & 1,507 & 62,376 & 34.7 & 41.3 & 97.69\% & Music & 301 & 1,206 \\
    \bottomrule 
  \end{tabular}
  \end{adjustbox}
\end{table}


\subsection{Baselines}
\label{sec:baselines}
To analyse the effectiveness of our provider fairness method, we compared its performance to that of two traditional models (WMF \cite{de2022modelling,tran2018regularizing} and PF \cite{mnih2007probabilistic}) and two neural network-based recommendation models (NeuMF \cite{xue2017deep,zou2020towards} and VAECF \cite{yalcin8empirical}). In addition to assessing accuracy, our target is to examine the method's ability to meet other important beyond-accuracy criteria, such as diversity, novelty, and serendipity. By including these beyond-accuracy aspects in our evaluation, we were able to provide a more comprehensive analysis of providing provider fairness compared to the traditional and neural network-based models.
    

    

\subsection{Evaluation Metric}
\label{sec:metrics}
While accuracy is a crucial metric to assess the performance of recommender systems, it is vital to recognize that it is not the sole indicator of a system's effectiveness. Other essential factors to consider while evaluating a recommender system's performance include diversity, novelty, and exposure. Therefore, to assess the performance of recommender systems and the proposed provider bias strategy, we rely on accuracy-based metrics including \textbf{Precision} \cite{arif2022destinations}, \textbf{Recall} \cite{arif2022destinations,cremonesi2010performance,ashok2009debugadvisor}, \textbf{NDCG} \cite{han2017survey}, and beyond-accuracy metrics, namely,  \textbf{Diversity} \cite{han2017survey,faria2023evaluating}, \textbf{Novelty} \cite{kaminskas2016diversity}, \textbf{Coverage} \cite{ge2010beyond}, \textbf{Serendipity} \cite{ge2010beyond}, \textbf{Personalisation} \cite{ge2010beyond}, and \textbf{Item Exposure} \cite{naghiaei2022cpfair}.

A diverse and novel recommendation system can help users discover new and unexpected items that they may not have considered otherwise. Additionally, serendipity measures the system's ability to offer unique and pleasant surprises to users beyond their usual preferences.

\section{Result}
\label{sec:result}

\begin{table*}
\centering
\caption{The recommendation performance of our re-ranking method and corresponding baselines on Epinion and BookCrossing datasets. The evaluation metrics here are calculated based on the top-10 predictions in the test set. Our best results are highlighted in \textbf{bold}. Rel$_{Short}$ and Rel$_{long}$ values denote the number of relevant recommended short-head and long-tail items. Fairness-unaware algorithm is shown with the notation $N$ and our proposed re-ranking method is denoted with $P$. }
\label{tbl:explicit_results}
\begin{adjustbox}{width=380pt}
\begin{tabular}{llllllllllllllll}
\toprule
\multirow{2}{*}{Model} & \multirow{2}{*}{Type} & \multicolumn{3}{c}{Accuracy} && \multicolumn{5}{c}{Beyond-Accuracy} && \multicolumn{4}{c}{Item Exposure} \\
\cmidrule{3-5}\cmidrule{7-11}\cmidrule{13-16}
& & NDCG & Pre & Rec && Nov. & Div. & Cov. & Per. & Ser. && Short. & Rel$_{Short}$ & Long. & Rel$_{Long}$ \\
\midrule
\multicolumn{16}{c}{\textbf{Epinion}} \\ \hline
PF & N & 0.0321 & 0.0312 & 0.0445 && 4.9602 & 0.9153 & 0.5073 & 0.9601 & 0.9388 && 22,169 & 8,637 & 4,601 & 3,595 \\ 
PF & P & \textbf{0.0323} & \textbf{0.0314} & \textbf{0.0448} && \textbf{4.9915} & \textbf{0.9161} & \textbf{0.5112} & \textbf{0.9613} & \textbf{0.9391} && \textbf{8,316} & \textbf{7,768} & \textbf{18,454} &	\textbf{4,774} \\ \hline
WMF & N & \textbf{0.0235} & \textbf{0.0225} & \textbf{0.0317} && 4.9484 & 0.9274 & 0.2913 & 0.9366 & 0.9441 && 25,225 & 6,566 & 1,545 & 1,276 \\ 
WMF & P & 0.0225 & 0.0215 & 0.0296 && \textbf{5.1762} & \textbf{0.9319} & \textbf{0.3587} & \textbf{0.9457} & \textbf{0.9467} && \textbf{20,846} & \textbf{6,316} & \textbf{5,924} & \textbf{2,110} \\ \hline
NeuMF & N & \textbf{0.0451} & \textbf{0.0406} & \textbf{0.0574} && 3.8923 & 0.8708 & 0.1214 & 0.7714 & 0.9257 && 26,533 & 6,276 & 237 & 291 \\ 
NeuMF & P & 0.0442 & 0.0392 & 0.0556 && \textbf{3.9874} & \textbf{0.8743} & \textbf{0.1340} & \textbf{0.7853} & \textbf{0.9267} && \textbf{24,443} & \textbf{6,128} & \textbf{2,327} & \textbf{591} \\ \hline
VAECF & N & 0.0444 & 0.0410 & 0.0597 && 4.3809 & 0.8794 & 0.2893 & 0.9038 & 0.9266 && 24,707 & 7,998 & 2,063 & 1,850 \\ 
VAECF & P & \textbf{0.0445} & 0.0410 & \textbf{0.0600} && \textbf{4.4043} & \textbf{0.8804} & \textbf{0.2947} & \textbf{0.9052} & \textbf{0.9269 }&& \textbf{15,676} & \textbf{7,522} & \textbf{11,094} & \textbf{2,622} \\
\midrule
\multicolumn{16}{c}{\textbf{BookCrossing}} \\ \hline
PF & N & 0.0108 & 0.0106 & \textbf{0.0276} && 5.8712 & 0.9620 & 0.8940 & 0.9780 & 0.9716 && 5,858 & 1,554 & 5,502 & 1,715 \\ 
PF & P & \textbf{0.0111} & \textbf{0.0107} & 0.0271 && \textbf{5.9188} & \textbf{0.9624} & \textbf{0.8989} & \textbf{0.9794} & \textbf{0.9721} && \textbf{337} & \textbf{794} & \textbf{11,023} & \textbf{1,952} \\ \hline
WMF & N & 0.0062 & 0.0059 & 0.0158 && 6.6007 & 0.9733 & 0.9715 & \textbf{0.9802} & 0.9770 && 1,968 & 1,323 & 9,392 & 1,998 \\ 
WMF & P & \textbf{0.0062} & \textbf{0.006} & \textbf{0.0161} && \textbf{6.6111} & \textbf{0.9735} & \textbf{0.9725} & 0.9801 & \textbf{0.9771} && 0 & 0 & \textbf{11,360} & \textbf{2,061} \\ \hline
NeuMF & N & 0.0165 & 0.0147 & 0.0380 && 4.5143 & 0.9422 & 0.0343 & 0.2026 & 0.9671 && 11,360 & 411 & 0 &	0 \\ 
NeuMF & P & 0.0165 & 0.0147 & 0.0380 && 4.5143 & 0.9422 & 0.0343 & 0.2026 & 0.9671 && 11,360 & 411 & 0 &	0 \\ \hline
VAECF & N & \textbf{0.0211} & \textbf{0.0189} & 0.0473 && 5.1439 & 0.9440 & 0.4004 & 0.9165 & 0.9630 && 9,287 & 1,446 & 2,073 &	648 \\ 
VAECF & P & 0.0190 & 0.0177 & \textbf{0.0475} && \textbf{5.5079} & \textbf{0.9430} & \textbf{0.4524} & \textbf{0.9338} & \textbf{0.9647} && \textbf{5,672} & \textbf{1,339} & \textbf{5,688} & \textbf{826} \\ \hline
\bottomrule
\end{tabular}
\end{adjustbox}
\end{table*}


In this section, we provide a thorough examination of the beyond-accuracy metric in the fairness-aware recommendation performance in comparison to fairness-unaware baseline models. Our analysis seeks to understand the trade-offs 
among different item groups in terms of beyond-accuracy objectives.

\subsection{Fairness-aware Algorithm Effectiveness.}
Trends in Tables~\ref{tbl:explicit_results} and \ref{tb2:explicit_results} show that the fairness-aware algorithm effectively increases the exposure of long-tail items while reducing the prominence of short-head items. This shift allows a broader range of items to gain exposure, providing a more equitable recommendation environment for providers. The redistribution of item exposure has several implications. First, it benefits providers by offering a fairer distribution of exposure, enabling long-tail items to compete with more popular items and potentially gain traction among users. For instance, as you see in Table~\ref{tbl:explicit_results} for Epinion the total number of short-head and long-tail recommended items using PF baseline has changed from $(22169, 4601)$ to $(8316, 18454)$, respectively. Second, it contributes to the overall diversity and novelty of the recommended items, enhancing the user experience and encouraging them to explore new content while keeping the relevant popular items. 
For example, for the case of PF and on Epinion dataset in Table~\ref{tbl:explicit_results}, one can see improvement along all beyond-accuracy metrics.
Furthermore, the number of relevant short-head and long-tail recommended items (denoted as Rel$_{short}$ and Rel$_{long}$) changes from $(8637, 3595)$ to $(7768, 4774)$ indicating the model capability of choosing relevant popular items while giving more exposure to relevant long-tail items. Similar trends can be observed in other baselines and datasets. This equitable distribution aligns with the goals of our fairness-aware algorithm, as it seeks to strike a balance between promoting fairness and maintaining the quality of recommendations.




\begin{table*}
\centering
\caption{The recommendation performance of our re-ranking method and corresponding baselines on Gowalla and Last.fm datasets. The evaluation metrics here are calculated based on the top-10 predictions in the test set. Our best results are highlighted in \textbf{bold}. Rel$_{Short}$ and Rel$_{long}$ values denote the number of relevant recommended short-head and long-tail items. Fairness-unaware algorithm is shown with the notation $N$ and our proposed re-ranking method is denoted with $P$.}
\label{tb2:explicit_results}
\begin{adjustbox}{width=380pt}
\begin{tabular}{llllllllllllllll}
\toprule
\multirow{2}{*}{Model} & \multirow{2}{*}{Type} & \multicolumn{3}{c}{Accuracy} && \multicolumn{5}{c}{Beyond-Accuracy} && \multicolumn{4}{c}{Item Exposure} \\
\cmidrule{3-5}\cmidrule{7-11}\cmidrule{13-16}
& & NDCG & Pre & Rec && Nov. & Div. & Cov. & Per. & Ser. && Short. & Rel$_{Short}$ & Long. & Rel$_{Long}$\\
\midrule
\multicolumn{16}{c}{\textbf{Gowalla}} \\ \hline
PF & N & 0.0592 & 0.0558 & 0.0568 && 4.0587 & 0.8462 & 0.6098 & 0.9615 & 0.8730 && 7,939 & 4,884 &	3,361 &	3,233 \\ 
PF & P & \textbf{0.0592} & \textbf{0.0559} & \textbf{0.0571} && \textbf{4.0868} & \textbf{0.8467} & \textbf{0.6165} & \textbf{0.9631} & \textbf{0.8733} && \textbf{2,085} & \textbf{4,192} & \textbf{9,215} &	\textbf{4,375} \\ \hline
WMF & N & 0.0338 & 0.0347 & 0.0368 && 4.5012 & 0.8862 & 0.4617 & 0.9563 & 0.8937 && 6,526 & 3,777 &	4,774 &	2,173 \\ 
WMF & P & 0.0339 & 0.0346 & 0.0365 && \textbf{4.5163} & \textbf{0.8871} & \textbf{0.4626} & \textbf{0.9567} & \textbf{0.8942} && 40 & 386 &	\textbf{1,1260} &	\textbf{2,618} \\ \hline
NeuMF & N & \textbf{0.0563} & \textbf{0.0528} & \textbf{0.0536} && 3.3409 & \textbf{0.8127} & 0.1463 & 0.8815 & 0.8649 && 10,447 & 3,451 &	853 &	577 \\ 
NeuMF & P & 0.0509 & 0.0485 & 0.0470 && \textbf{3.6512} & 0.8093 & \textbf{0.1615} & \textbf{0.8901} & \textbf{0.8657} && \textbf{7,361} & \textbf{3,309} &	\textbf{3,939} &	\textbf{877} \\ \hline
VAECF & N & \textbf{0.0652} & \textbf{0.0625} & \textbf{0.0673} && 3.8219 & 0.8092 & 0.4079 & 0.9543 & 0.8579 && 8,025 & 4,360 &	3,275 &	2,724 \\ 
VAECF & P & 0.0569 & 0.0548 & 0.0560 && \textbf{4.3186} & \textbf{0.8145} & \textbf{0.4407} & \textbf{0.9552} & \textbf{0.8682} && \textbf{3,299} & \textbf{3,739} &	\textbf{8,001} &	\textbf{3,304} \\
\midrule
\multicolumn{16}{c}{\textbf{Last.fm}} \\ \hline
PF & N & \textbf{0.0372} & 0.0350 & 0.0469 && 5.0905 & 0.9201 & 0.6682 & 0.9775 & 0.9213 && 12,480 & 6,978 &	5,490 &	2,967 \\ 
PF & P & 0.0371 & 0.0350 & \textbf{0.0470} && \textbf{5.1089} & \textbf{0.9206} & \textbf{0.6709} & \textbf{0.9779} & \textbf{0.9215} && \textbf{10,724} & \textbf{6,939} & \textbf{7,246} & \textbf{3,133} \\ \hline
WMF & N & \textbf{0.0319} & \textbf{0.0301} & \textbf{0.0434} && 5.5358 & 0.9294 & 0.7804 & 0.9723 & 0.9257 && 8,000 & 6,267 &	9,970 &	3,272 \\ 
WMF & P & 0.0279 & 0.0262 & 0.0380 && \textbf{5.8797} & \textbf{0.9420} & \textbf{0.7963} & \textbf{0.9725} & \textbf{0.9328} && 0 & 0 &	\textbf{17,970} &	\textbf{3,688} \\ \hline
NeuMF & N & \textbf{0.0415} & \textbf{0.0390} & \textbf{0.0519} && 3.7975 & 0.8647 & 0.0916 & 0.8802 & 0.9034 && 17,863 & 4,866 &	107 &	67 \\ 
NeuMF & P & 0.0387 & 0.0366 & 0.0488 && \textbf{3.8812} & \textbf{0.8694} & \textbf{0.0962} & \textbf{0.8843} & \textbf{0.9049} && \textbf{16,665} & \textbf{4,835} & \textbf{1,305} & \textbf{184} \\ \hline
VAECF & N & 0.0555 & 0.0514 & 0.0725 && 4.6164 & 0.8769 & 0.4280 & 0.9661 & 0.8985 && 13,998 & 6,746 &	3,972 &	1,920 \\ 
VAECF & P & \textbf{0.0556} & \textbf{0.0517} & \textbf{0.0732} && \textbf{4.6635} & \textbf{0.8794} & \textbf{0.4326} & \textbf{0.9669} & \textbf{0.8996} && \textbf{6,786} & \textbf{6,354} & \textbf{11,184} &	\textbf{2,406} \\ \hline
\bottomrule
\end{tabular}
\end{adjustbox}
\end{table*}

\subsection{Beyond-accuracy Metrics Influence.}
\begin{figure*}
  \centering
  \subfloat[NDCG]
    {\includegraphics[scale=0.22]{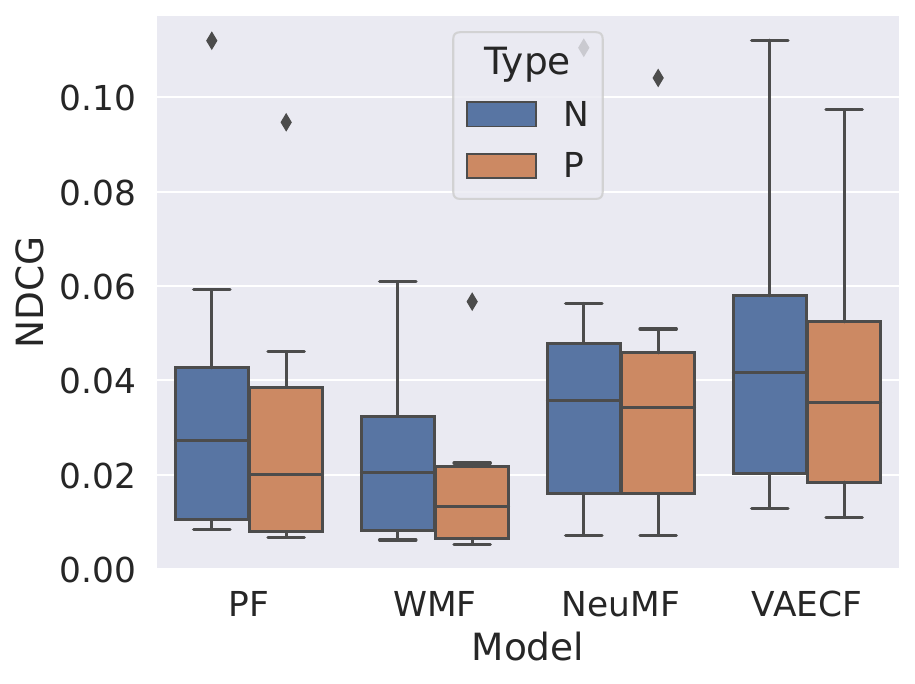}
    \label{fig:boxplot_NDCG}}
  \hfill
  \subfloat[Pre]
    {\includegraphics[scale=0.22]{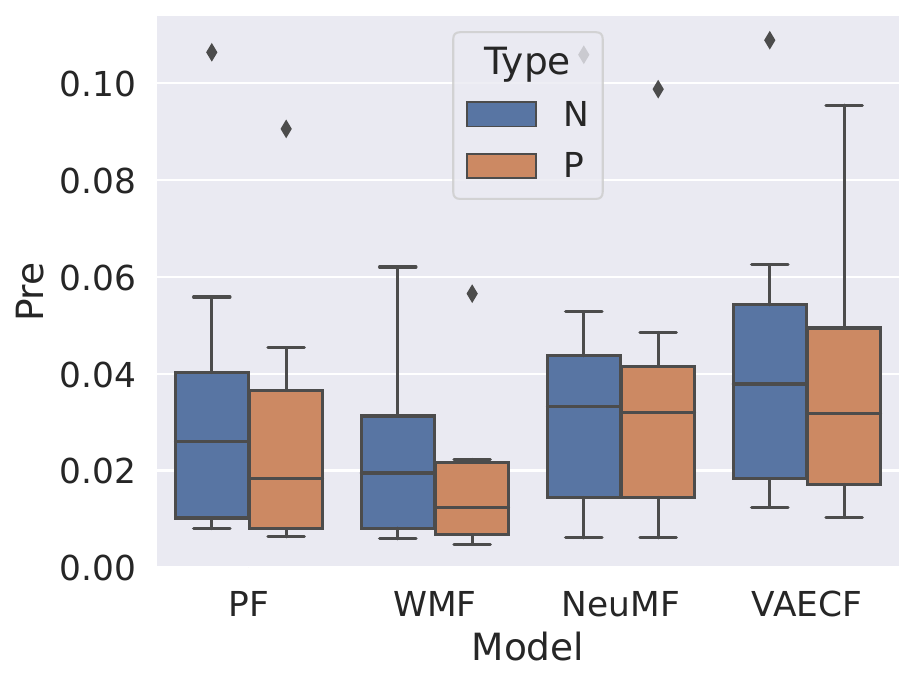}
    \label{fig:boxplot_Pre}}
  \hfill
  \subfloat[Rec]
    {\includegraphics[scale=0.22]{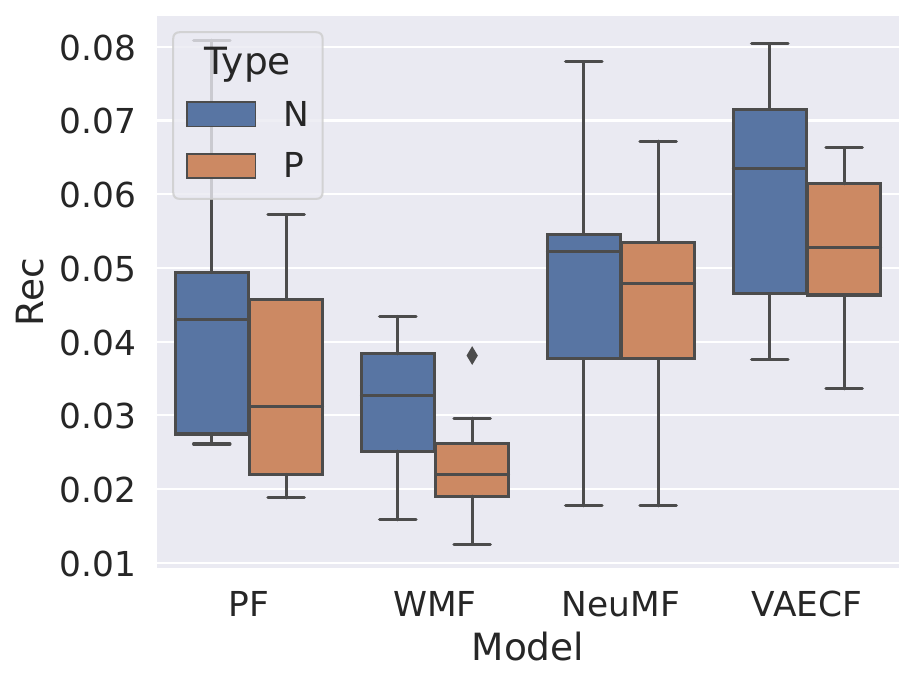}
    \label{fig:boxplot_Rec}}
  \hfill
  \subfloat[Nov]
    {\includegraphics[scale=0.22]{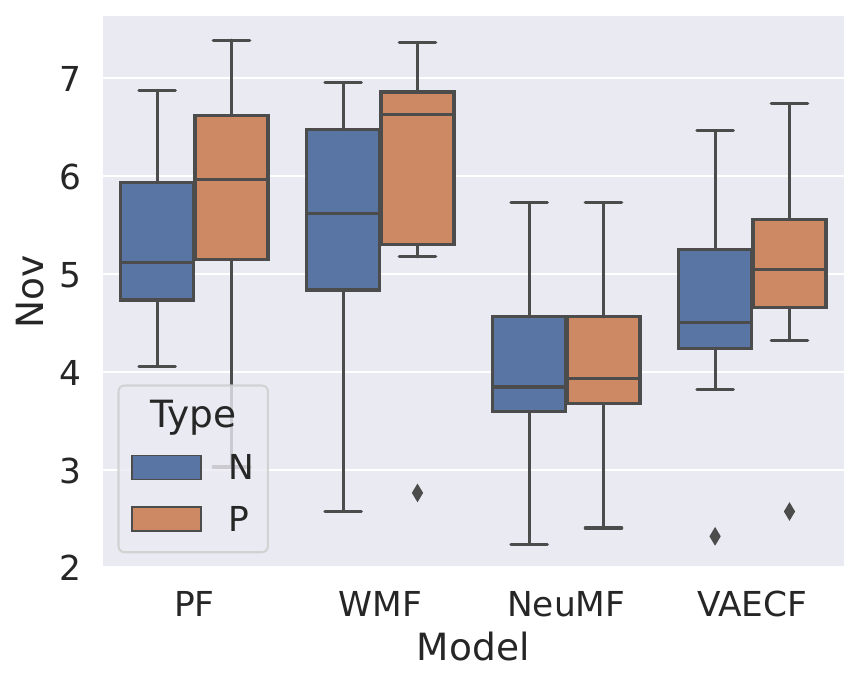}
    \label{fig:boxplot_Nov}}
\hfill
  \subfloat[Div]
    {\includegraphics[scale=0.22]{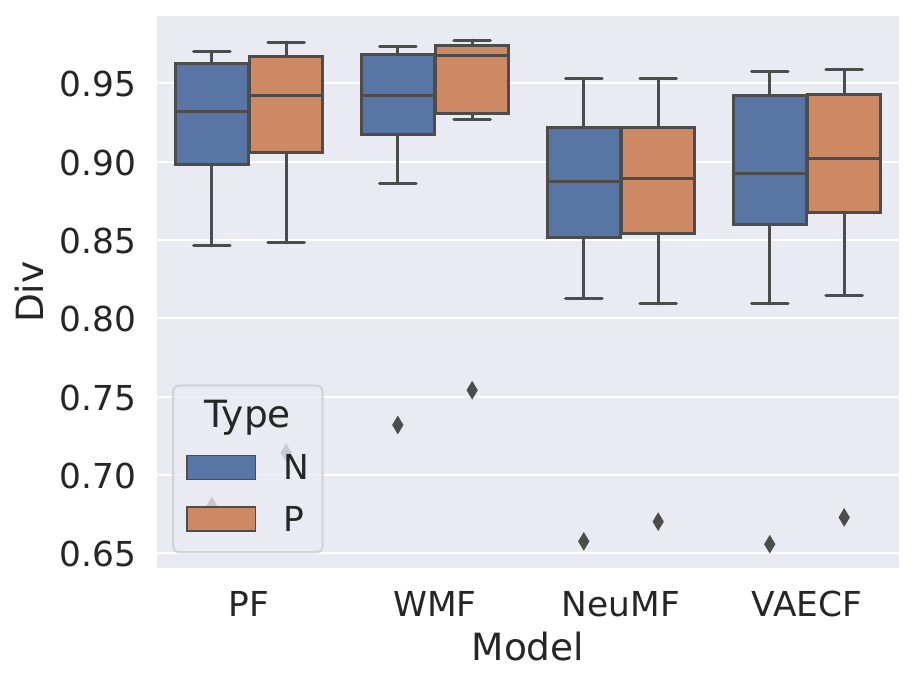}
    \label{fig:boxplot_Div}}
    \hfill
  \subfloat[Cov]
    {\includegraphics[scale=0.22]{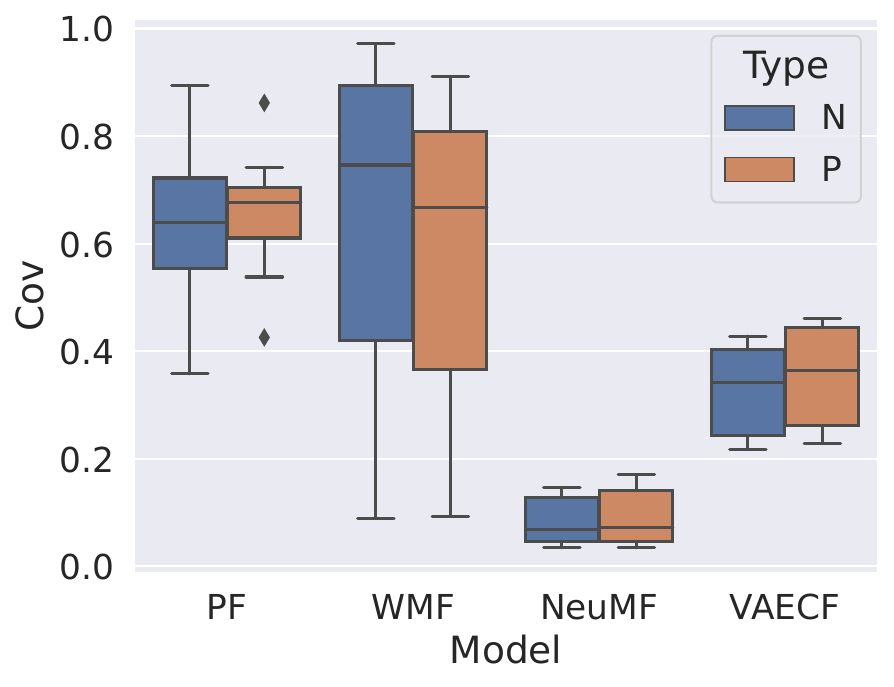}
    \label{fig:boxplot_Cov}}
    \hfill
  \subfloat[Per]
    {\includegraphics[scale=0.22]{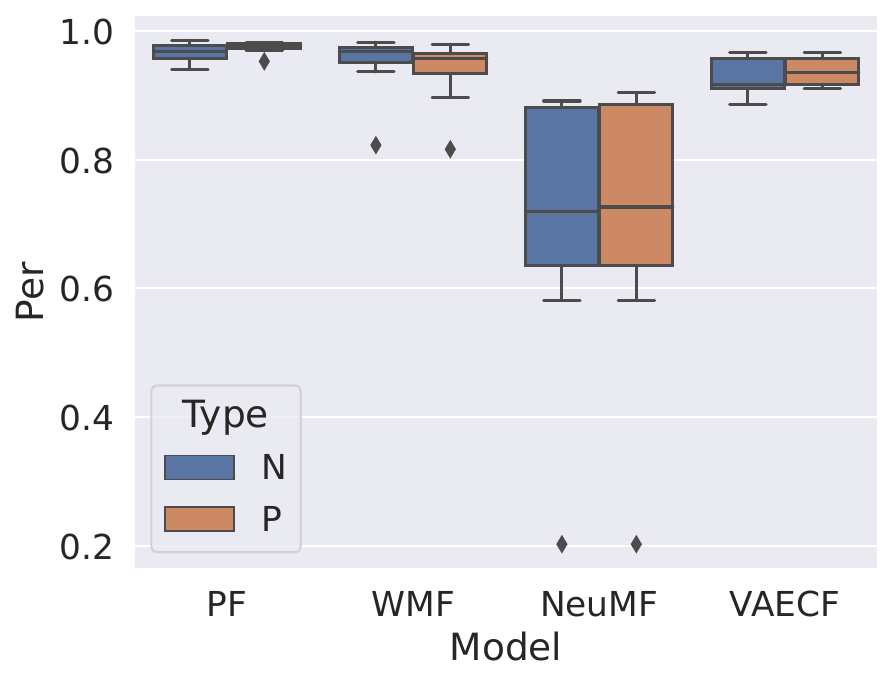}
    \label{fig:boxplot_Per}}
    \hfill
  \subfloat[Ser]
    {\includegraphics[scale=0.22]{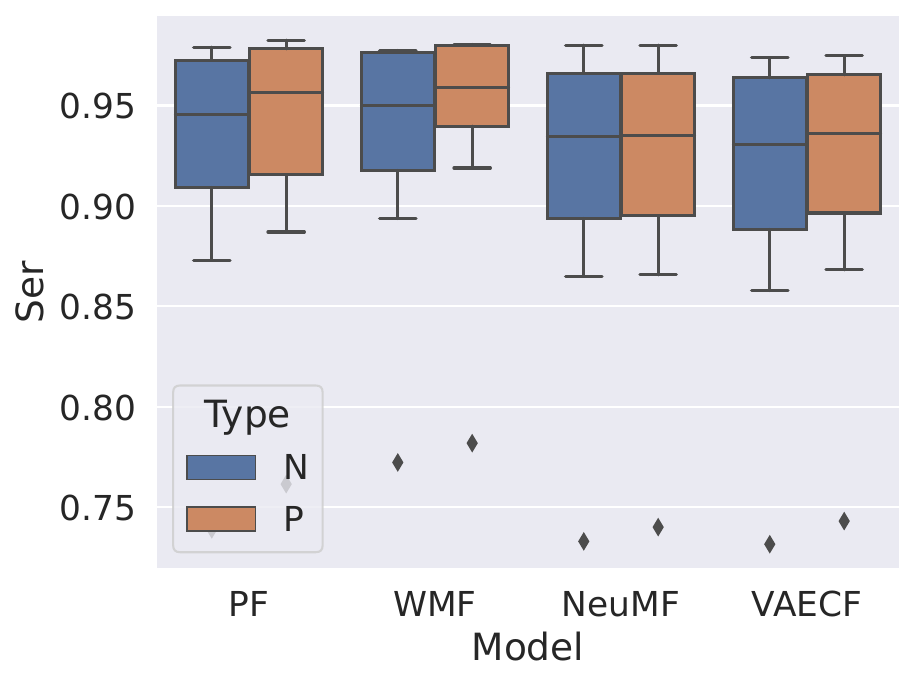}
    \label{fig:boxplot_Ser}}
  \caption{Comparative performance of fairness-unaware and fairness-aware models across key evaluation metrics. Each box represent the variation among all the datasets.}
\label{fig:item_unfairness_rec}
\end{figure*}

According to Fig.~\ref{fig:item_unfairness_rec}, the implementation of our proposed algorithm has led to a consistent upward trend in beyond-accuracy metrics across all models and datasets. For example, as can be seen in Table~\ref{tbl:explicit_results}, we observe notable improvements in novelty, diversity, coverage, and serendipity for all models in Epinion. For instance, the WMF model's novelty increased from 4.94 to 5.17, while diversity in the VAECF model rose from 0.87 to 0.88. Similar positive trends are observed in the BookCrossing dataset.
Overall, the novelty has the most improvement in the Last.fm dataset, while diversity and coverage have the best improvement in the Epinion dataset. This can be due to the underlying data characteristics of the datasets. However, among the models, the coverage in the NeuMF model has relatively low values, although it has increased in the fairness-aware model, its result is insignificant compared to other models. The personalisation and serendipity also have fewer changes compared to novelty and diversity, but they improved more in Epinion.

\subsection{Dataset and Baseline Model Dependency.}
Our analysis reveals dependencies on both the dataset and the baseline model used when implementing the fairness-aware algorithm. The BookCrossing dataset, for example, exhibited more significant improvements across various criteria than other datasets, such as Gowalla. This dataset dependency can be attributed to the unique data structure and characteristics within each dataset, which can affect the outcomes of the research and alter the values obtained from the implementation of the models. On the other hand, the baseline model also plays a crucial role in the effectiveness of the fairness-aware algorithm. The VAECF model, in particular, exhibits the most substantial enhancement in beyond-accuracy criteria due to its ability to capture complex and non-linear relationships between user preferences and item features through its variational autoencoder framework. Conversely, the NeuMF model shows comparatively less improvement in beyond-accuracy criteria, possibly due to its reliance on a combination of matrix factorization and multi-layer perceptron techniques. In conclusion, the performance differences among models can be ascribed to their distinct underlying mechanisms, which consider user preferences, item features, and intricate interactions. Nonetheless, applying the fairness-aware reranking to all models consistently improved beyond-accuracy metrics, with the extent of this improvement varying notably among the models.

\section{Discussion}
In this section we summarize the answers we found to the research questions we listed in Section \ref{sec:intro}.

\subsection{RQ1: Impact on exposure of long-tail items and provider Fairness.} 
Our findings demonstrate that the proposed post-processing re-ranking optimization framework effectively increased the exposure of long-tail items, resulting in a more diverse and equitable distribution of recommendations. On average, the number of recommended and relevant recommended long-tail items for fairness-unaware and fairness-aware models are $(3565.8, 2116.6)$ and $(8765, 2573.1)$, respectively. This enhancement in exposure subsequently contributed to improving overall provider fairness.

\subsection{RQ2: Consequences of improving provider fairness on other aspects of recommendation quality.} 
By improving provider fairness, we observed positive effects on other aspects of recommendation quality, such as diversity, serendipity, and novelty. This suggests that promoting fairness can enhance the user experience by offering more diverse and unexpected recommendations, catering to a wider range of user interests and preferences. However, it is important to strike a balance between increasing fairness and maintaining the quality of recommendations. Overemphasizing fairness could potentially result in recommendations that do not align with users' interests or needs, leading to dissatisfaction and reduced system efficiency. Thus, it is crucial to exercise moderation when suggesting unpopular items, ensuring that popular items are still included in the recommendations.

\subsection{RQ3: Trade-offs between provider fairness and recommendation quality across different algorithms and datasets.} Our analysis indicates that the trade-offs between provider fairness and recommendation quality manifest differently across various recommendation algorithms and real-world datasets. The nature of the dataset, data structure, and data characteristics can affect the performance of fairness-aware algorithms and influence the balance between fairness and recommendation quality. The choice of recommendation algorithm also plays a role in managing this trade-off. While some algorithms may perform better in certain datasets, they may not yield the same results in others. As such, it is essential to consider the specific context and requirements of a recommendation system when selecting and implementing a fairness-aware algorithm.


\section{Conclusion}
\label{sec:conclusion}
In conclusion, this study presents a fairness-aware re-ranking algorithm designed to mitigate biases in recommendation systems, specifically addressing the overemphasis on popular items. By incorporating a comprehensive set of beyond-accuracy evaluation metrics, including novelty, diversity, coverage, and serendipity, we thoroughly analyze the impact of our fairness-aware approach on these metrics and their implications for item providers. The results demonstrate that our fairness-aware approach has a positive impact on these beyond-accuracy metrics, with only a minor reduction in recommendation accuracy. This indicates that the overall effectiveness of the system is not significantly compromised when provider is introduced in this way.
\bibliographystyle{ACM-Reference-Format}
\bibliography{sample-base}


\end{document}